\documentclass[12pt]{article}
\usepackage{amsmath,amssymb}

\usepackage{xcolor}
\begin{document}

\begin{center}

\section*{Towards  quantum turbulence theory:\\  A simple model with interaction of the vortex loops.}

{S.V. TALALOV}

{Department of Applied Mathematics, Togliatti State University, \\ 14 Belorusskaya Str.,
 Tolyatti, Samara Region, 445020 Russia.\\
svt\_19@mail.ru}

\end{center}

\begin{abstract}
This paper investigates quantized  thin  vortex rings with an internal structure.  The quantization scheme of this dynamical system is based on an earlier
 the approach proposed by the author.  
Both energy spectrum and circulation spectrum are calculated. Examples show that the set of permissible circulation values has a fractal structure.
The suggested model allows us  to describe  the system of isolated vortex  rings as well as  the vortex rings with  interaction.
Furthermore, the application to the quantum turbulence theory is discussed.  The general expression for the partition function of a turbulent flow is suggested.
\end{abstract}

{\bf keywords:}   quantum vortex rings,  quantum vortex interaction, quantum fluids turbulence.

\vspace{5mm}

 {\bf PACS numbers:}         47.32.C \quad 47.32.cb  \quad  47.27.-i

\vspace{5mm}

\section{ Introduction}	

~~~ The complexity of such a phenomenon as turbulence leads to the emergence of different approaches to its description. 
For example, any attempts to describe  turbulent motion of a fluid  using the Navier-Stokes equations lead to significant difficulties, even at the classical level.
This fact, in particular, stimulates the search for other approaches to the description of this phenomenon.
The description of such  motion at the quantum level presents an even more complex problem.
It is now an established fact that vortex structures play a primary role in the formation of turbulent flows of quantum fluids.
A large number of works are devoted to this issue. Without reviewing
  the literature on this topic, we will mention some of the works 
\cite{Feyn,Donn,Aarts,TsFuYu,Nemir,MuPoKr}.
{ It can be assumed that  investigation of simplified models of quantum turbulence will be no less useful than, for example, simplified models in the quantum field theory. 
For example,  such models could  provide some progress in calculating the thermodynamic characteristics of a turbulent flow (entropy, Gibbs free energy, etc.)}

In this paper, we propose a simple model of quantized vortices that demonstrates the following properties:
\begin{itemize}
\item Scale and Galilean invariance of the theory;
\item A broad spectrum of energy and circulation values.  In particular, the set of circulation values found has a fractal structure.
 In our opinion, this result is quite suitable for describing the random distribution of circulation in a turbulent flows;
\item The ability to describe the interaction of vortex loops,  as well as  the   reconnecting of such a loop and   resizing it.
\end{itemize}
Of course, the phenomenon of quantum turbulence is too complicated to be described completely in one paper. 
Here we  consider a simple model that allows us to calculate the permissible values of energy, circulation and certain other variables.
For example,  the suggested approach gives the explicit formulas for quantized fluid velocity in some points of  the fluid flow.
We also  describe the interaction of quantized vortex rings,  creation and annihilation of the vortices included.
Within the framework of our assumptions,   the proposed theory makes it possible to write a general expression for the  partition function  of a turbulent flow.
{ The author hopes that the subsequent development of the model, including the refinement of the resulting expression for the partition function,
 will be useful for some thermodynamic calculations.}

 As a starting point of our research, we consider  the special configurations of the closed vortex filaments with an internal  core structure. We suppose that the dynamics of  such objects  is restricted  by the local induction approximation.    
Under certain assumptions \cite{AlKuOk}, a vortex filament  $ {\boldsymbol{r}}(t ,s)$  with a nonzero flow inside the core is described by the equation:

\begin{eqnarray}
        \label{LIE_dim}
        \partial_t {\boldsymbol{r}} (t ,s) &  = &
        A \,\partial_s{\boldsymbol{r}}(t ,s)\times\partial_s^{\,2}{\boldsymbol{r}}(t ,s) + \nonumber\\
				~~ & + &  B\Bigl(\partial_s^{\,3}{\boldsymbol{r}}(t ,s) + 
        \frac{3}{2}\,\bigl\vert\, \partial_s^{\,2}{\boldsymbol{r}}(t ,s)\bigr\vert^{\,2}\partial_s{\boldsymbol{r}}(t ,s)\Bigr)\,.
				        \end{eqnarray}
								
We use    notations  $t$ and  $s$  for  the time and  the  natural   curve  parameter     correspondingly. 
  Coefficients $A$ and $B$  are  some dimensional coefficients  which depend on circulation $\Gamma$, the radius ${\sf a}$ of the vortex core 
and the components of the flow velocity in the core. Regarding the value ${\sf a}$, we consider it finite and small enough.
 In general, all these values  can vary from vortex to vortex in a turbulent flow.

The theory has three natural  dimensional constants  that are relevant to the physical system being described.
These constants are: the fluid's density $\varrho_0$, the speed of sound in this fluid $v_0$ and the natural scale length $R_0$:
$$R_0 \in  \{\,R\,:   R =   |\,{\boldsymbol r}_1 - {\boldsymbol r}_2|\,, \qquad {\boldsymbol r}_1, {\boldsymbol r}_2 \in V\, \}\,,$$
 where   symbol  $V$  denotes the domain where the investigated objects evolve. For example, the constant $R_0$  may be the radius of the pipe in which the fluid in question flows. 
 Despite the fact that  value  $\tilde\mu_0 =  \varrho_0 R_0^3$   is a natural parameter that determines the scale of the masses,
  we will  also   use  the  additional  mass   parameter $\mu_0 $.
	In our theory this parameter denotes
 the central charge for central extension of the Galilei group\footnote{The appearance of the extended Galilei group in the considered approach was discussed in the author's work \cite{Tal} in detail. } $\widetilde{\mathcal G}_3$. Therefore, we have an additional dimensionless parameter here:
   $\alpha_{\sf ph} = \mu_0/\tilde\mu_0$. We will clarify its role in our theory later.  	
Along with constants $\varrho_0$, $v_0$, $R_0$, we will use the auxiliary  constants $t_0 = R_0/v_0$ and   ${\cal E}_0  = \mu_0 v_0^2$.

The model under consideration allows us to consider a separate vortex ring as a particle with an internal degree of freedom. 
As a consequence,   it becomes possible
to  use the  standard tools of quantum many body  theory to describe the interaction of such rings.
For example,  processes of   creation and annihilation of the  vortices in a fluid flow
can be described.
Note that the suggested approach is much simpler 
than using field string theory which is usually discussed in this context.

\section{ Classical dynamics of a single vortex}

 ~~~ Let the symbol $R$ denotes the arbitrary positive constant with the dimension of length. 
Along with the ''physical''  vectors  ${\boldsymbol{r}}$,
we define here the projective vectors ${\boldsymbol{\large\textsf{r}}}   = {\boldsymbol{r}}/R$.
Further, to describe the considered vortex filament, we introduce the  dimensionless parameters
$\tau = t/t_0$ and $\xi = s/R$.  As a consequence, the Eq. (\ref{LIE_dim}) will be rewritten as

\begin{eqnarray}
        \label{LIE_str}
        \partial_\tau {\boldsymbol{\large\textsf{r}}}(\tau ,\xi)  & = &
        \beta_1 \Bigl(\partial_\xi{\boldsymbol{\large\textsf{r}}}(\tau ,\xi)\times\partial_\xi^{\,2}{\boldsymbol{\large\textsf{r}}}(\tau ,\xi)\Bigr)   + \nonumber \\
				~~ & + & \beta_2\Bigl(2\,\partial_\xi^{\,3}{\boldsymbol{\large\textsf{r}}}(\tau ,\xi) + 
        {3}\,\bigl\vert\, \partial_\xi^{\,2}{\boldsymbol{\large\textsf{r}}}(\tau ,\xi)\bigr\vert^{\,2}\partial_\xi{\boldsymbol{\large\textsf{r}}}(\tau ,\xi)\Bigr)\,.
				        \end{eqnarray}
								
The  values  $\beta_1$ and $\beta_2$  are dimensionless constants here.

Equation (\ref{LIE_str}) has a certain solution that is of interest to the proposed model. This solution is:
\begin{equation}
        \label{our_sol}
 \boldsymbol{\large\textsf{r}}(\tau ,\xi) = \Bigl(\, \frac{q_x}{R} +  \cos(\xi +\phi_0 +\beta_2\tau)\,, \frac{q_y}{R} +  \sin(\xi +\phi_0+\beta_2\tau  )\,, \frac{q_z}{R} + \beta_1 \tau \,\Bigr)\,, 
\end{equation}
where the angle   $\phi_0 \in [0, 2\pi)$ and the coordinates $q_x, q_y, q_z$   are some
 (time - independent) variables. 
This solution describes the vortex filament  in the shape of a circle with a radius $R$.   The filament  moves along the axis  $\boldsymbol{e} = {\boldsymbol{e}}_z$
with velocity $|\boldsymbol{u}_{v}| = \beta_1 R/t_0$ and  rotates with  the frequency  $\beta_2/t_0$.  The rotation  simulates  here  the flow $\Phi$  inside the  filament  core.
This flow is:
\begin{equation}
        \label{Flow_class}
				\Phi =  \beta_2\pi \varrho_0{\sf a}^2 v_0  \bigl(  R/R_0\bigr)\,. \nonumber
\end{equation}
We will further consider only such solutions of Eq. (\ref{LIE_str}). Thus, the set of possible vortex loops is reduced  to the rings of an arbitrary radius and some fluid flow in the core here.

In addition to Eq. (\ref{LIE_str}) that describes the evolution of the  
 curve ${\boldsymbol{r}}(\cdot,\xi)$, we postulate the standard hydrodynamic formula \cite{Batche}
for the momentum   ${\boldsymbol{p}}$:
	  
   \begin{equation}
        \label{p_and_m}
        {\boldsymbol{p}} = \frac{\varrho_0}{2 }\,\int\,\boldsymbol{r}\times\boldsymbol{\omega}(\boldsymbol{r})\,dV\,.
                 \end{equation}

The vector  $\boldsymbol{\omega}(\boldsymbol{r})$  stands for   vorticity. 
In our model 
$$ \boldsymbol{\omega}(\boldsymbol{r})  =  \boldsymbol{\omega}_1(\boldsymbol{r}) +    \boldsymbol{\omega}_2(\boldsymbol{r})\,,$$
where the vorticity  $\boldsymbol{\omega}_1(\boldsymbol{r})$ is due to the fluid   rotation  around the  filament 
and the vorticity  $\boldsymbol{\omega}_2(\boldsymbol{r})$ is due to  the fluid flow in the filament core.
Let us consider these summands separately. In a cylindrical coordinates $(\rho, \varphi, z)$, the  fluid flow  velosity $\boldsymbol{u}$   in the filament core is described  by the formula
$$  \boldsymbol{u}(\boldsymbol{r})  = const\times\delta(\rho - R)\delta(z) \boldsymbol{e}_\varphi(\varphi)\,.$$
Therefore, 
$$ \boldsymbol{r}\times\boldsymbol{\omega}_2(\boldsymbol{r}) = \boldsymbol{r}\times \bigl(\nabla \times  \boldsymbol{u}(\boldsymbol{r})\bigr) = C(\rho, z)\boldsymbol{e}_\varphi(\varphi)\,.$$
Because the equality   $\int_0^{2\pi} \boldsymbol{e}_\varphi(\varphi) d \varphi = 0$  holds, the value $\boldsymbol{\omega}_2(\boldsymbol{r})$ does not contribute to the integral Eq. (\ref{p_and_m}).
Regarding  the first summand  $\boldsymbol{\omega}_1(\boldsymbol{r})$,  
  the vorticity of the  closed vortex filament  is calculated by means of the formula (see, for example, Ref. \cite{AlKuOk})
   \begin{equation}
        \label{vort_w}
     \boldsymbol{\omega}_1(\boldsymbol{r}) =  \Gamma
                  \int\limits_{0}^{2\pi}\,\delta(\boldsymbol{r} - \boldsymbol{r}(\xi))\partial_\xi{\boldsymbol{r}}(\xi)d\xi\,,
       \end{equation}  
where the symbol 	$\Gamma$ stands for circulation.

Taking into account the formulas   Eqs.  (\ref{p_and_m}) and   (\ref{vort_w}),   we deduce  the following  expression for the canonical momentum:
	\begin{equation}
        \label{impuls_def1}
                 {\boldsymbol{p}}  =   \frac{\varrho_0\Gamma {R}^{\,2}}{2}                    
                 \iint\limits_{0}^{2\pi}  \left[\, {\xi - \eta}\,\right]\,\partial_\eta{\boldsymbol{\large\textsf{r}}}(\tau ,\eta)
								\times\partial_\xi{\boldsymbol{\large\textsf{r}}}(\tau ,\xi)d\xi  d\eta\,.
      \end{equation} 
The notation $[\,x\,]$ means the integer part of the number $x/{2\pi}$ here.

For our solution Eq. (\ref{our_sol}) the integral in r.h.s. of the  formula Eq. (\ref{impuls_def1}) is easily calculated. Therefore,
\begin{equation}
        \label{impuls_our1}
                {\boldsymbol{p}}  =   \pi \varrho_0 {R}^2 \Gamma  {\boldsymbol{e}} \,,\qquad   |{\boldsymbol{e}}|  = 1\,,                
                       \end{equation} 
where  constant unit vector  ${\boldsymbol{e}}$  defines the axis of the rotating ring Eq. (\ref{our_sol}).

Earlier,   a new approach to the Hamiltonian description and quantization of a single vortex loop  was proposed by the author\cite{Tal}.
In this paper, we modify the suggested approach for our purposes.  Moreover we perform the reduction of the considered dynamical
 system  to the finite number    of degrees of  freedom.

As can be seen from the formulas Eqs. (\ref{our_sol})  and   (\ref{p_and_m}), the natural variables that parametrize   our dynamical system are  variables
\begin{equation}
        \label{var_in}
 \boldsymbol{q} =  (\,q_x\,,\quad q_y\,,\quad q_z\,)\,,\qquad R\,,\qquad 
\phi = \phi(\tau) = \phi_0 +\beta_2\tau \,,\quad \Gamma\,, \qquad {\boldsymbol{e}}\,,  
\end{equation} 
where   $|{\boldsymbol{e}}|  = 1$. 
 It must be noted that we describe   the  vortex ring as an  abstract closed curve which evolves in accordance with Eq. (\ref{LIE_str}). 
The   variable  as ''velocity of the fluid''  will not be needed for this purpose,
 and still we intend to take into account the dynamics of the surrounding fluid in a minimal way:
			we declare the value $\Gamma$ as a dynamic variable, in addition to   variables $\boldsymbol{q}$,  $R$, $\phi$  and  ${\boldsymbol{e}}$.

The variables Eqs. (\ref{var_in}), for example,    enable us to determine the core  radius  ${\sf a}$   in our model. Indeed, if ${\sf a} \simeq 0$, the following formula for
the vortex ring velocity $\boldsymbol{u}_{v}$  takes place \cite{Saffm}:
$$ |\boldsymbol{u}_{v}|  \simeq  \frac{\Gamma}{4\pi R}\ln\frac{8R}{{\sf a}}\,.$$
In our model $|\boldsymbol{u}_{v}| = \beta_1 R/t_0$; therefore , the following expression   is true  for the value ${\sf a}$:
\begin{equation}
        \label{core_radius}
		{\sf a} \simeq 8 R \exp\left(-  \frac{4\pi\beta_1 R^2}{t_0 \Gamma}  \right)\,.		
\end{equation}

As  the next step,  we  replace the  natural set of variables Eqs. (\ref{var_in}) by another one which is more convenient. Let us define the variables
$$  \varpi = \frac{R}{R_0}\cos(\phi_0 +\beta_2\tau)\,, \qquad   \chi = \frac{R}{R_0}\sin(\phi_0 +\beta_2\tau)\,.$$
Dynamical equations for these variables are canonical Hamiltonian equations for a harmonic oscillator:
\begin{eqnarray}
\partial_\tau  \varpi & = &  -  \beta_2 \chi\,, \nonumber\\
  \partial_\tau  \chi  & = &   \beta_2 \varpi\,. \nonumber
\end{eqnarray}

The formula Eq. (\ref{impuls_our1}) is than rewritten as follows:
\begin{equation}
        \label{impuls_our2}
                 {\boldsymbol{p}}  =   \pi  \varrho_0 {R_0}^2 \Gamma \bigl(\varpi^2  +  \chi^2\bigr){\boldsymbol{e}} \,,\qquad   |{\boldsymbol{e}}|  = 1\,.                
                       \end{equation}

Apparently, the set of the variables ${\boldsymbol{p}}$,  $\boldsymbol{q}$, $\varpi$  and  $\chi$ adequately describes  our dynamical system Eq. (\ref{our_sol})
 as a structured  $3D$ particle with an
internal degree of the freedom. 
The formula Eqs. (\ref{impuls_our2}) together with the definition of the variables  $\varpi$  and  $\chi$ provides one-to-one correspondence  between the set Eqs.  (\ref{var_in})
and the new set (${\boldsymbol{p}}$,  $\boldsymbol{q}$, $\varpi$, $\chi$).  Note that the variables $\varpi$ and $\chi$  are invariants under Galilean and scale transformations of space $E_3$.
It might be appropriate to remember   Lord Kelvin's old idea  \cite{Thom}
about interpretation  of vortices  as some  structured particles.     This idea   is still being discussed at the present time \cite{Moff}.

The next step in the development of our model  is the Hamiltonian description of the  considered dynamical system.
Pursuant to the Dirac's  prescriptions about the primacy of the Hamiltonian structure,
  we define such structure  axiomatically here.  The relevant definitions are given below.
   \begin{itemize}
  \item Phase space ${\mathcal H} =  {\mathcal H}_{pq}  \times  {\mathcal H}_b   $. The space $ {\mathcal H}_{pq}$ is the phase space of a $3D$  free structureless    particle.
	It is  parametrized by the variables 
   ${\boldsymbol{q}}$ and  ${\boldsymbol{p}}$.  The space    $ {\mathcal H}_b$  is a phase space for one-dimensional harmonic oscillator.
		 \item Poisson structure:
  \begin{eqnarray}
  \{p_i\,,q_j\} & = & \delta_{ij}\,,\qquad i,j = x,y,z\,, \nonumber \\
  \label{ja_jb}
  \{ \varpi, \chi\} & = & \frac{1}{{\cal E}_0 t_0 }
  \end{eqnarray}
  	All other brackets  vanish. 
\item Hamiltonian 
\begin{equation}
        \label{hamilt_1}
				H = \frac{\boldsymbol{p}^2}{2 \mu_0} +  \frac{\beta_2{\cal E}_0}{2}\Bigl(\varpi^2  + \chi^2  \Bigr)\,.
	\end{equation}
	\end{itemize}
One of the main questions here is how we  can  describe the energy of the vortex rings under consideration?
It is a well-known fact that the  canonical formula \cite{Saffm}
       \begin{equation}
  \label{can_energy}
       {\mathcal E} = \frac{1}{8\pi}\,\iint
       \frac{\boldsymbol{\omega}(\boldsymbol{r})\boldsymbol{\omega}(\boldsymbol{r}^{\prime})}{|\,\boldsymbol{r} - \boldsymbol{r}^{\prime}|}\,dVdV^{\prime} =
			\frac{{\Gamma}^{\,2}}{8\pi}\iint 
       \frac{\partial_\xi{\boldsymbol{r}}(\xi)\partial_\xi{\boldsymbol{r}}(\xi^{\prime})}{|\,{\boldsymbol{r}}(\xi) -  {\boldsymbol{r}}(\xi^{\prime}) |}\,d\xi d\xi^{\prime}		
			\nonumber
       \end{equation}
			leads to the unsatisfactory result for thin filaments.  Indeed, the integral in this formula   diverges  for the filament with the core radius  ${\sf a} \to 0$. 
			The standard approach to solving this problem is to use various regularization methods. 				In our case, where the value ${\sf a} $ depends on  dynamic variables of the theory [see Eq. (\ref{core_radius})], such a procedure would look somewhat ambiguous. 
			There exists yet another method  proposed in Ref. \cite{Tal}: the energy of an arbitrary closed vortex filament is considered from a group-theoretic point of view there.
			This approach is based on the fact that 	
	           Lee algebra of the  group $\widetilde{\mathcal G}_3$ has three Cazimir functions:  
	
	 $$ {\hat C}_1 = \mu_0 {\hat I}\,,\quad 
  {\hat C}_2 = \left({\hat M}_i  - \sum_{k,j=x,y,z}\epsilon_{ijk}{\hat P}_j {\hat B}_k\right)^2 
  \quad {\hat C}_3 = \hat H -  \frac{1}{2\mu_0}\sum_{i=x,y,z}{\hat P}_i^{\,2}\,,$$                         
       where        ${\hat I}$ is the unit operator,     ${\hat M}_i$,   $\hat H$,  ${\hat P}_i$         and  ${\hat B}_i$  ($i = x,y,z$)
        are the respective generators of rotations, time and space translations and Galilean boosts. 
		Traditionally, the function  ${\hat C}_3 $  can be interpreted as  an  ''internal energy of the particle''. 
		In our case it is a naturall postulate that 		      			
       $${ C}_3  			=   {\beta_2{\cal E}_0} |\,b\,|^2\,, \qquad b = \frac{ \chi   + i \varpi}{\sqrt{2}}    	\,.$$ 
Therefore, the identification   ${\mathcal E} = H$ is justified in our approach.

Finally, the proposed approach enables us to consider the vortex  as a point particle with coordinates $\boldsymbol{q}$  and momentum $\boldsymbol{p}$.
 Each such particle has  an internal degree of freedom  which is described by oscillator variables $\chi$   and  $\varpi$. 
These variables define the  radius $R$ of the vortex and the flow in the vortex core.
 We assume that the coordinates $\boldsymbol{q}$  are the coordinates of the center of the vortex ring.
Therefore,  taking into account the definition of circulation, the following expression for the fluid velocity  $\boldsymbol{u}_f$  takes place  in this point:
\begin{equation}
	\label{vel_fl}
	\boldsymbol{u}_f  =  \boldsymbol{u}_v  +    \frac{\Gamma}{2\pi R} \boldsymbol{e} = \left(\frac{\beta_1 R}{t_0} +    \frac{\Gamma}{2\pi R} \right) \boldsymbol{e}\,, 
	\qquad   \boldsymbol{e} =  \frac{\boldsymbol{p}}{|\boldsymbol{p}|}\,.
	\end{equation}

\section{Quantization}

	~~~ The constructed Hamiltonian structure     defines the way for quantization of the vortex  ring being studied.
										First, we must define a Hilbert space $\boldsymbol{H}_1$   of the quantum states of our dynamical system.  The structure of the phase space ${\mathcal H}$ lead to the 
					following natural structure of the space $\boldsymbol{H}_1$:
					\begin{equation}
	\label{space_quant}
	\boldsymbol{H}_1  =  \boldsymbol{H}_{pq} \otimes   \boldsymbol{H}_b\,,
	\end{equation}
			where the symbol   $\boldsymbol{H}_{pq}$  denotes the Hilbert space  of a free structureless $3D$ particle  
						(space $L^2({\sf R}_3)$ for example) and the  symbol $\boldsymbol{H}_b $ 
						denotes the  Hilbert space of the quantum states for the  harmonic oscillator.		
						The creation and annihilation operators $\hat{b}^+$,   $\hat{b}$ as well as the standard orthonormal basis $  |\,n\rangle$
							  in the space 	$\boldsymbol{H}_b $  are defined  by well-known formulas 												
				$$ [\,\hat{b}, \hat{b}^+] = \hat{I}_b\,, \qquad \hat{b}|\,0_b\rangle = 0  \,,
				\qquad  |\,n\rangle   =  \frac{1}{\sqrt{n!}} (\hat{b}^+)^n    |\,0_b\rangle   \qquad    |\,0_b\rangle \in    \boldsymbol{H}_b           \,,$$   
				where the operator   $\hat{I}_b$  is a unit operator in the space    $\boldsymbol{H}_b $.
				
	Let us quantize our theory.  For certainty, we will consider  the  case when 		$\boldsymbol{H}_{pq} = L^2({\sf R}_3)$	 and 	$\boldsymbol{H}_b = L^2({\sf R})$ only.
	According to the classical quantization scheme, 
	we must  construct the function $A \to \hat{A}$, where symbol $A$ denotes some classical variable and symbol $\hat{A}$ denotes some operator in the space $\boldsymbol{H}_1$.
	For fundamental Hamiltonian variables , the relation 
	$$  [\hat{A},\hat{B}]  = -i\hbar \widehat{\{A,B\}}\,$$
		must be satisfied.   		This equality can possess  some  ''anomalous terms''  if the ''observables'' $A$, $B$ are the functions of the fundamental variables.
					These terms depend  on the ordering rule of  non-commuting operators.   We will not discuss these issues here \cite{Berezin}.
					Thus, our postulate of quantization is as follows:
					$$ q_{x,y,z} \to  q_{x,y,z}\otimes \,\hat{I}_b  \,,\qquad  p_{x,y,z} \to - i\hbar\frac{\partial}{\partial q_{x,y,z}} \otimes \,\hat{I}_b \,,\qquad
					b \to \sqrt{\frac{\hbar}{t_0{\mathcal E}_0}}\, (\hat{I}_{pq} \otimes\,\hat{b})\,, $$
					where    operator $\hat{I}_{pq}$  is a unit operator in the space  $\boldsymbol{H}_{pq}$.  					
We will not  subsequently write the constructions $(\dots  \otimes \,\hat{I}_b)$  and  $ (\hat{I}_{pq} \otimes\,\dots)$ explicitly, hoping that this will not lead to misunderstandings.
In accordance with our quantization postulates, the Hamiltonian is defined by the operator
	\begin{equation}
        \label{hamilt_q}
			\hat{	H} = \frac{\hbar^2}{2 \mu_0} \Delta   +  \frac{\beta_2 \hbar}{ t_0}  \Bigl( \hat{b}^+ \hat{b} + \frac{1}{2} \Bigr)\,.
	\end{equation}
	
		To find  possible values of circulation $\Gamma$, let's square the  Eq. (\ref{impuls_our2}).
After quantization, we have the following equation:
\begin{equation}
        \label{for_Gamma1}
\left(\hbar^2\Delta  + \pi^2 \sigma_{\sf ph}^4\varrho_0^2 \Gamma^2 R_0^4 \Bigl[\,(\hat{b}^+)^2\hat{b}^2  + 2\hat{b}^+ \hat{b} +\frac{1}{4}\,\Bigr]\right)|\Psi\rangle =0\,, 
\qquad  |\Psi\rangle \in  \boldsymbol{H}\,,
	\end{equation}			
			where $	\sigma_{\sf ph} = \sqrt{{\hbar}/{t_0{\mathcal E}_0}}$. We will discuss 	this constant later.	
The eigenvalues  ${\mathcal E}$  of the operator  $\hat{	H}$ and   specific values of the quantity  $\Gamma$ depend on the domain  $V$ in which the motion of the vortex ring in question occurs.
Before considering a specific example, we would like to say a few words  about the possible values of quantized circulation  $\Gamma  = \Gamma_n$  in a turbulent flow.
In our opinion:
\begin{itemize}
\item[$\star$] The conventional  formula  
\begin{equation}
        \label{Circ_post}
				\Gamma_n \equiv  \oint_\gamma {\bf u}(\ell)d{\boldsymbol\ell}   =\frac{ n \hbar}{\mu}\,\qquad  n = 0,1,2,\dots,		
\end{equation}	
   { can be refined }  for  a turbulence. Apparently, the set of values of the quantized quantity $\Gamma_n$ is significantly wider here than the natural series.
Of course, in the simplest cases, the formula  Eq. (\ref{Circ_post})   remains valid,
possibly with some correcting terms (see author's work \cite{Tal_PoF}).
Note that the known experimental measurements of the magnitude of $\Gamma$ were made for special single vortices,
and not in a turbulent flow.  {  An overview of the results on this issue is given Ref. \cite{Donn}. }
In Ref.  \cite{MuPoKr} , the formula Eq. (\ref{Circ_post})  was confirmed by the results of numerical modeling { in the framework of the Gross-Pitaevskii model.} 
 Let us note that these works predict
the ''large'' peaks in the integer values and certain ''small'' peaks at the non-integer values,   { which} was explained by errors.
  From the author's point of view, 
 { alternative models are quite appropriate for such a complex phenomenon as a quantum turbulence.}    
\item[$\star$] The rule Eq. (\ref{Circ_post}) is usually postulated.  
 As   has been repeatedly stated in the literature, such quantization rules are similar to the quantization rules in the old Bohr quantum theory.
The author believes that quantum  values $\Gamma_n$  should be deduced  from the general postulates of quantum theory, and not postulated separately.
\item[$\star$] As for the arguments that rule Eq. (\ref{Circ_post}) is a consequence of the unambiguity of the wave function of certain
 quasi-particles\footnote{In a two-fluid model, which we are not considering here.}, we will make the following remark.  
We consider the       fluid  medium  where the closed vortex loops with  ''zero''  thickness are present.   This medium   can be considered as a
 realization  of multi-connected space.  Thus, any quasi-particle here can possess the fractional statistics. 
In this case, the wave function of a quasi-particle can receive a phase multiplier when moving along a closed path around a vortex filament.
Therefore, the condition Eq. (\ref{Circ_post})  may not hold  even for a single vortex in general. Here we should mention
 the Ref. \cite{Protog}, where the anyon super-conductivity was investigated (as is well known, this phenomenon is similar to super-fluidity).
\end{itemize}

Let us consider the following example. We assume that the fluid moves  in the domain $V$, which is determined by the following boundary conditions:
$$ x^2 + y^2  \le R_0^{\,2}\,, \qquad    z \in [0, 2\pi R_1]\,\, (mod\,2\pi R_1)\,, \quad R_1 = const\,. $$
This domain  models a   round   tube with a radius $R_0$  in the shape of a torus of radius $R_1 >> R_0$.

{ The effects on the boundaries of the domain $V$ can be modeled by conditions on the wave function $\Psi({\boldsymbol r})  \in \boldsymbol{H}_{pq}$  
 on the surface  $ x^2 + y^2  = R_0^{\,2}$.  It is clear that $\Psi({\boldsymbol r}) = \psi(x,y)\psi(z)$ here. Therefore, we can consider the following conditions:
$$ c_1 \frac{\partial \psi(x,y)}{\partial{\boldsymbol n} }\Big\vert_{x,y \in S}   + c_2\psi(x,y) \Big\vert_{x,y \in S}   =    0\,,$$
where  ${\boldsymbol n}$ is the normal vector for the circle $S: \{x,y: x^2 + y^2  = R_0^{\,2}\}$. 
 Let us consider the simplest case when $c_1 =0$ and $c_2 = const$.}
It is known that eigenvalues $-\lambda^2$  of the Laplace operator in  { this case} 
 will be  values  $\lambda_{n,k}^2 =   (\zeta^{(n)}_k/R_0)^2$,
where  quantities $\zeta^{(n)}_k$, $k = 1,2,\dots$  stand for  zeros of the Bessel function $J_n(\rho)$.
Thus, the eigenvalues $-\lambda^2$  of the Laplace operator $\Delta$ in the  domain $V$  take the following values:
$$  \lambda_{m,\ell,k}^2  =  \left(\frac{m}{2 R_1}\right)^2 +  \left(\frac{\zeta^{(\ell)}_k}{R_0}\right)^2\,, \qquad  m,\ell = 0,1,2,\dots\,\quad    k = 1,2,\dots\,.$$
As a result , we find the following formulas for energy  $\mathcal E$  and circulation $\Gamma$:
\begin{eqnarray}
        \label{energy_t}
\mathcal E_{n,m,\ell,k} & = & \frac{\hbar^2 \lambda_{m,\ell,k}^2}{2\mu_0} + \frac{\beta_2 \hbar }{ t_0} \Bigl(n +\frac{1}{2}\,\Bigr)  \,,\\[2mm]
        \label{circ_t}
				\Gamma_{n,m,\ell,k} & = & \pm\frac{2\hbar R_0\lambda_{m,\ell,k} }{\sigma_{\sf ph}^2\tilde\mu_0 (2n + 1)} =
				\frac{\pm 2 \hbar  }{\sigma_{\sf ph}^2\tilde\mu_0 (2n + 1)}\sqrt{   \left(\frac{{m} R_0}{2 R_1}\right)^2 +  \left({ \zeta^{(\ell)}_k  }\right)^2}\,,
\end{eqnarray}
where natural numbers $n, m,\ell = 0,1,2,\dots\,$  and number  $k = 1,2,\dots$.  As is well known, the asymptotic behavior of values  $\zeta^{(\ell)}_k$  for large values
$\ell$ and $k$ will be   $\zeta^{(\ell)}_k  \simeq (3\pi/4) + (\pi/2)\ell + \pi k$.
Therefore, any asymptotics $s \to \infty$, where  number $s$ is the number $m$ or $\ell$ or $k$ gives the formula Eq. (\ref{Circ_post}) for
circulation $\Gamma_{n,m,\ell,k}$ if other quantum numbers are fixed.

Let's establish the properties of the set $\{\Gamma_{n,m,\ell,k}\}$.
\begin{itemize}
\item The set $\{\Gamma_{n,m,\ell,k}\}$ has a fractal structure.
First, we verify  the   property of self-similarity. It is more convenient to
  take the set
$$\{\Gamma^2\} =  \{\Gamma_{n,m,\ell,k}^{\,2}\,;\,\,  n, m,\ell = 0,1,2,\dots\,, k = 1,2,\dots \}$$
for this purpose.
Indeed,
$$\{\Gamma^2\} = \bigcup_{n,m} {\hat D}_n {\hat T}_m \{\Upsilon\}\,,$$
where the set $\{\Upsilon\}$ is the set of  points ${2\pi \zeta^{(\ell)}_k R_1 }/{R_0}$ on the real axis, symbol ${\hat T}_m$ stands for translation
$x \to x + const\times m^2$ and symbol ${\hat D}_n$  stands for dilatation $x \to x\times ({\hbar  }/\mu_0 (2n + 1))^2$.
To calculate the fractal dimension, let 's write the set  $\{\Gamma_{n,m,\ell,k}\}$  in the form
$$\{\Gamma_{n,m,\ell,k}\} =  \bigcup_{m,\ell,k} \{X_{m,\ell,k}\}\,,   $$
where the sets $\{X_{m,\ell,k}\}$ are the sequences $X_n = \Gamma_{n,m,\ell,k}$,  $n =0,1,2,\dots$,  where  numbers $m,\ell,k$ are fixed.
These sequences have asymptotic behavior as $const/n$ when $n \to \infty$.
Then, the distance  $\delta$ between neighboring elements $X_n$  and $X_{n+1}$  at $n \to \infty$  is equal to $\delta(n) = 1/n^2$.
Applying the standard formula for calculating the fractal dimension  ${\mathcal D}$, we find
$$ {\mathcal D}_{X}  =  \lim_{n \to \infty}\frac{\ln n}{\ln (1/ \delta(n))} = \frac{1}{2}\,.$$
This result is well known for the fractal dimension of the natural series.
Therefore, the set of the circulation values in our model demonstrates  fractal properties. 
In our opinion, such a structure of the set  $\{\Gamma_{n,m,\ell,k}\}$  is more suitable for describing a turbulent flow 
 then the ''regular'' structure due to the formula Eq. (\ref{Circ_post}).
\item The set $\{\Gamma_{n,m,\ell,k}\}$ is bounded.  Indeed, the following inequality takes place  for the  physical reasons:
$$  \frac{\hbar^2 \lambda_{m,\ell,k}^2}{2\mu_0} < {\mathcal E}_{max}\,,$$
where the constant  ${\mathcal E}_{max}$ is some maximal energy in the considered flow.  Consequently,
\begin{equation}
\label{circ_bound}
\Gamma < 2 R_0  \frac{\alpha_{\sf ph}}{\sigma_{\sf ph}^2} \sqrt{\frac{2{\mathcal E}_{max} }{\mu_0}}\,.
\end{equation}
\end{itemize}

The pure quantum states   $|n;m,\ell,k\rangle \in \boldsymbol{H}_1$ that correspond to the values Eqs.  (\ref{energy_t}) and  (\ref{circ_t}) are written as follows:
\begin{equation}
        \label{pure_st}
|n;m,\ell,k\rangle  =  |m,\ell,k\rangle |n\rangle\,,\qquad |m,\ell,k\rangle \in \boldsymbol{H}_{pq}\,, \quad |n\rangle \in \boldsymbol{H}_a\,,
\end{equation}
where the notation $|m,\ell,k\rangle$ was used for the eigenvectors of Laplace operator.

Similarly, we can calculate the values of $\mathcal E$ and $\Gamma$ in an arbitrary domain $V$. Indeed, let the numbers $-\lambda_{[s]}^2$  be the eigenvalues of the Laplace operator
 in the domain $V$, where the notation  $[s]$   means some multi-index (such as  the complex index $\{m,\ell,k\}$ in example above). Corresponding formulas
for the values of $\mathcal E$ and $\Gamma$ will be similar to the formulas Eqs. (\ref{energy_t}) and      (\ref{circ_t}).
  Having excluded the value $-\lambda_{[s]}^2$ from these formulas, we find a connection between energy and circulation:   
			\begin{equation}
        \label{E_Gamma}							
							{\mathcal E}_{[s],n}  =  \frac{1}{2}{\tilde\mu_0 \Gamma_{[s],n}^2}
								\sigma_{\sf ph}^4\alpha_{\sf ph}								
								\left( \frac{n + 1/2}{R_0} \right)^2	+
								\beta_2 \hbar\, v_0 \left( \frac{n + 1/2}{R_0} \right)\,.
		\end{equation}

The radius $R$ of considered vortices is also quantized. Indeed, 
$$ R^2 \to \hat{R}^2      =  \frac{\hbar R_0^2}{t_0{\mathcal E}_0}\Bigl( \hat{b}^+ \hat{b} + \frac{1}{2} \Bigr)\,.$$
Therefore, we have following values for the radius $R = R_n$:
\begin{equation}
        \label{R_quant}
 R_n =\sigma_{\!\sf ph}  R_0\sqrt{    n + \frac{1}{2}}\,, \qquad     n = 0,1, \dots,    \,.  
\end{equation}
The dimensionless constant   $\sigma_{\!\sf ph} =  \sqrt{  {\hbar} / {\mu_0 v_0 R_0}}\,$  also appeared here.  
This constant, in addition to the previously introduced dimensionless constant $\alpha_{\sf ph}$, naturally appears in the quantum version of the  considered dynamical system.
These constants depend on both specific fluid and domain $V$.
For example, if the value $\mu_0$ equals   the  ${}^4{\rm He}$  mass, the value $v_0 \simeq 3.4\, m/c$ (the sound speed in the liquid Helium) and the ''pipe radius''  $R_0 \simeq 0.03\, m$, 
we have  values   $\alpha_{\!\sf ph}  \simeq  2 \cdot10^{-27}$ and 
 $\sigma_{\!\sf ph} \simeq 10^{-3}$ for these constants.

\section{Description of the many-vortex systems}

~~~ Here we describe  the vortex loop as some point-like particle with the internal degree of the freedom. 
This approach  gives   possibilities for studying  many-vortex flows. 
First, we consider the fluid flow which contains      $N$ non-interacting vortices   numbered by some multi-index $[n]$.
 For the sake of clarity, we will assume that the fluid is in the volume $V$, which was introduced in the previous section.
We will also make the following assumptions:
\begin{enumerate}
\item   As is well-known, some space averaging  is needed  for the description of a turbulent flow in  a concrete physical system.  
We assume that    the number of vortices  $N$ is sufficiently large, so that the space averaging volume  $\delta V$  contains a large number of the vortex loops with centers $\boldsymbol{q}_{[s]}$;
\item  The unequalities $R_n < l_a/2$ take place for every vortex of radius $R_n$ where the value  $l_a \simeq \sqrt[3]{V/N}$  is   average distance    between the vortex centers;
\item Let the  value  $l_1$ be  the distance in the fluid flow such that 
correlation for any parameters   in the points $\boldsymbol{q}_1$ and  $\boldsymbol{q}_2$   is absent  if $|\boldsymbol{q}_1  - \boldsymbol{q}_2| > l_1$. 
We suppose that the unequality $l_a  \ge l_1$ takes place.
\end{enumerate}
As mentioned earlier, we suppose that coordinates  $\boldsymbol{q}_{[k]}$  of our ''structured particle'' coincide with the center of the vortex ring.
 What is the quantized fluid velocity $\boldsymbol{u}_{[s]}$   in the point $\boldsymbol{q}_{[s]}$?  This value can be calculated as follows:
$$  \boldsymbol{u}_{[s]}  =  \boldsymbol{u}_{v} +  \frac{\Gamma_{n,m,\ell,k}}{2\pi R_n}\boldsymbol{e}\,,\qquad    |\boldsymbol{e}| = 1\,, $$
where the velocity  $\boldsymbol{u}_{v} = (\beta_1 R/t_0) \boldsymbol{e} $ of the vortex ring is defined in accordance with formula Eq. (\ref{our_sol}).
Therefore, the formulas Eqs.  (\ref{vel_fl}), (\ref{circ_t}) and (\ref{R_quant}) that were deduced earlier, give the following expression for the quantized fluid velocity in the point $\boldsymbol{q}_{[s]}$:
\begin{equation}
        \label{u_flow}
   \boldsymbol{u}_{[s]}   = 	\sigma_{\!\sf ph} v_0 	\left[ 	- \beta_1  \sqrt{ n + \frac{1}{2}} +
			\frac{1}{2(n +1/2)^{3/2}}\sqrt{   \left(\frac{{m} R_0}{2\pi R_1}\right)^2 +  \left({ \zeta^{(\ell)}_k  }\right)^2}\,\right]\boldsymbol{e} \,, 		 
\end{equation}

where the numbers $n,m,\ell,k$ are random natural numbers and the  vector $\boldsymbol{e}$ is a random  unit vector.
In accordance with the assumptions made, formula Eq. (\ref{u_flow})  models a quasi-random velocity distribution in  a turbulent flow.

The suggested  theory allows us to calculate the
partition function  ${\mathcal Z}$ for any concrete domain $V$. 
In the simplest case, when there is no interaction between vortices, the ${\mathcal Z}$ function can be written out explicitly.
 Taking into account the example from the previous section, 
we can write the following expression:
$$  {\mathcal Z}  = N\sum_{n,m,\ell,k}  \exp\left(- \frac{\mathcal E_{n,m,\ell,k} }{{\sf k}_B{\sf T}}\right)\,, $$
where  value ${\sf T}$ is the  temperature,  the constant ${\sf k}_B$ is  Boltzmann constant 
 and  the  energy levels  ${\mathcal E_{n,m,\ell,k} }$ have been determined  earlier by the  formula Eq. (\ref{energy_t}).

Of course, a system of non-interacting vortices is too unsatisfactory an approximation to describe a turbulent flow.
To get closer to reality, we must consider the interaction of vortices. The proposed theory allows us to do this.
Indeed, we can apply  the standard formalism  of  the   many-body systems  theory here.
Let us introduce     the  $N$-vortex space      $\boldsymbol{H}_N$:
$$ \boldsymbol{H}_N =   \underbrace{\boldsymbol{H}_1 \otimes  \dots \otimes \boldsymbol{H}_1}_{N}  \equiv 
{\mathfrak H}_{pq}^N  \otimes {\mathfrak H}_b^N\,,$$
where
$$ {\mathfrak H}_{pq}^N = 
 \underbrace{\boldsymbol{H}_{pq} \otimes  \dots \otimes \boldsymbol{H}_{pq}}_{N} \,,\qquad
{\mathfrak H}_b^N =  
  \underbrace{\boldsymbol{H}_b \otimes  \dots \otimes \boldsymbol{H}_b}_{N} \,.$$
In Dirac notation, any vector  $|\Phi^N\rangle \in  \boldsymbol{H}_N$ takes the form ($N \ge 1$)
\begin{eqnarray}
\label{vect_Phi}
|\Phi^N\rangle & = & \sum_{n_1,\dots,n_N} \int\,\cdots\int d\boldsymbol{p}_1\dots\boldsymbol{p}_N f^N(\boldsymbol{p}_1,\dots,\boldsymbol{p}_N)\times \nonumber\\
~& \times & \varphi^N_{n_1,\dots,n_N}|\boldsymbol{p}_1\rangle \dots|\boldsymbol{p}_N\rangle  | n_1\rangle\dots|n_N\rangle\,,
\end{eqnarray}
where the vectors  $|\boldsymbol{p}_j\rangle$ are corresponding eigenvectors of the operators $ \hat{\boldsymbol{p}}_j$.

 The Fock  space 
$${\mathfrak H}  =  \bigoplus_{N=0}^{\infty}\boldsymbol{H}_N  \subset   {\mathfrak H}_{pq} \otimes   {\mathfrak H}_{int} \,,    
        \qquad  \boldsymbol{H}_0 = |\,0_{pq}\rangle\otimes|\,0_b\rangle   =  {\sf C}   \,,   $$ 
is defined in a standard way.  
Here we have introduced the notation  
$${\mathfrak H}_{pq} = \bigoplus_{N=0}^{\infty} {\mathfrak H}_{pq}^N \,, \qquad 
{\mathfrak H}_{int} = \bigoplus_{N=0}^{\infty} {\mathfrak H}_b^N\,.  $$
Next, we consider the ”covering” space ${\mathfrak H}_{pq} \otimes   {\mathfrak H}_{int}$ only.
The creation and annihilation operators  ${\hat a}^+_{pq}(\boldsymbol{p})$,  ${\hat a}_{pq}(\boldsymbol{p})$ and  ${\hat a}^+_{int}(n)$,
${\hat a}_{int}(n)$ act in the space ${\mathfrak H}$ as ${\hat a}^+_{pq}(\boldsymbol{p})\otimes I_{int}$ and so on. They are defined in a standard way.
 For example, let us suppose  that  vectors $\Phi_b   \in {\mathfrak H}_{int}  $  take the form (vector in the form of a string):
$$\Phi_b   = \bigl(\varphi^0\,, \varphi^1_{n_1}\,,\dots, \varphi^N_{n_1,\dots,n_N}\,, \dots  \bigr)\,.$$
Then the definition of operators  ${\hat a}_{int}(n)$  and  ${\hat a}^+_{int}(n)$     will be as follows   ($\varphi^0 = 1$)
\begin{eqnarray}
\bigl({\hat a}_{int}(n)\varphi^N\bigr)_{n_1,\dots,n_{N-1}} & = &  \sqrt{N} \varphi^N_{n_1,\dots,n_{N-1},n}\,,   \qquad  \bigl({\hat a}_{int}(n)\varphi^0\bigr) =0\,, \nonumber\\
\bigl({\hat a}^+_{int}(n)\varphi^N\bigr)_{n_1,\dots,n_{N+1}} & = & \frac{1}{\sqrt{N+1}} \sum_{j=1}^{N+1} \delta_{n n_j}\,\varphi^N_{n_1,\dots,\not{n_j},\dots n_{N+1}}\,,\nonumber
\end{eqnarray}
where the symbol $\not{\!\!n_j}$ as well as the symbol $n_0$ both  mean the absence of the corresponding number. In addition, the symbol $n$ without any subscript is not a summation index
 in the formula Eq. (\ref{vect_Phi}). As usual, the constructions $(\dots\otimes I_a) $  and   $(I_{pq}\otimes \dots) $ will not be explicitly written out.
We will also consider the operators
$$ \hat{\boldsymbol{\mathfrak a}}^+(\boldsymbol{p};n) = {\hat a}^+_{pq}(\boldsymbol{p})\otimes {\hat a}^+_{int}(n)  \,, \qquad
\hat{\boldsymbol{\mathfrak a}}(\boldsymbol{p};n) =  {\hat a}_{pq}(\boldsymbol{p})\otimes {\hat a}_{int}(n) \,$$
which act  in the space ${\mathfrak H}$. 

Thus, the suggested method allows us to describe the processes of creation and annihilation of closed  vortex rings of a variable radius.
Indeed, let us consider the Hamiltonian
\begin{equation}
\label{ham_q_full}
\hat{H} =   \hat{H}_0 +   \hat{U}\,, 
\end{equation}
where
\begin{equation}
\label{ham_q_free}
\hat{H}_0 =   \frac{1}{2\mu_0}\int \boldsymbol{p}^2  {\hat a}^+_{pq}(\boldsymbol{p}){\hat a}_{pq}(\boldsymbol{p}) +
\frac{\beta_2 \hbar}{t_0} \sum_{n=0}^\infty \biggl(n + \frac{1}{2}  \biggr) {\hat a}^+_{int}(n){\hat a}_{int}(n)\,. 
\end{equation}
This operator has continuous  spectrum:
\begin{equation}
\label{spec_H0}
\hat{H}_0 \Phi^N_{\mathcal E} =  {\mathcal E} \Phi^N_{\mathcal E}\,, \nonumber
\end{equation}
where the eigenvalues ${\mathcal E}$  are positive numbers.
Let's find the eigenvectors $\Phi^N_{\mathcal E} \in \boldsymbol{H}_N $ of the operator  $\hat{H}_0$.
Introducing the designation
$$\Phi^N  = \bigl(0\,, 0\,,\dots, f^N(\boldsymbol{p}_1,\dots,\boldsymbol{p}_N)\varphi^N_{n_1,\dots,n_N}\,, 0\,,  \dots  \bigr)\,$$
and performing the direct calculations, we find for the vector $\Phi^N$:
\begin{equation}
\label{H0PhiN}
\hat{H}_0 \Phi^N \equiv  {\mathcal E}(\boldsymbol{p}_1,\dots  \boldsymbol{p}_N; n_1,\dots  n_N) \Phi^N\,,
\end{equation}
where
\begin{equation}
\label{eigen_1}
{\mathcal E}(\boldsymbol{p}_1,\dots  \boldsymbol{p}_N; n_1,\dots  n_N)  = 
  \frac{1}{2\mu_0} \sum_{j=1}^N \boldsymbol{p}_j^2   +  \frac{\beta_2 \hbar}{t_0} \sum_{j=1}^N \biggl(n_j + \frac{1}{2}  \biggr)\,.
\end{equation}
Therefore,
\begin{eqnarray}
\label{egenvect_H0}
|\Phi^N_{\mathcal E}\rangle & = & \sum_{n_1,\dots,n_N} \int\,\cdots\int d\boldsymbol{p}_1\dots\boldsymbol{p}_N 
f^N_{\mathcal E}(\boldsymbol{p}_1,\dots,\boldsymbol{p}_N; n_1,\dots  n_N)\times \nonumber\\
~& \times & \varphi^N_{n_1,\dots,n_N}|\boldsymbol{p}_1\rangle \dots|\boldsymbol{p}_N\rangle  | n_1\rangle\dots|n_N\rangle\,,
\end{eqnarray}
where the function $f^N_{\mathcal E}$ is proportional to the Dirac $\delta$-function:
$$f^N_{\mathcal E}(\boldsymbol{p}_1,\dots,\boldsymbol{p}_N; n_1,\dots  n_N) = const\times\delta\Bigl( {\mathcal E} - {\mathcal E}(\boldsymbol{p}_1,\dots  \boldsymbol{p}_N; n_1,\dots  n_N) \Bigr)\,.$$
Thus, the vector $|\Phi^N_{\mathcal E}\rangle$ is some entangled state of states  of   $N$ 
   vortices with momenta $\boldsymbol{p}_1,\dots,\boldsymbol{p}_N$ and radii ${R}_i$, $i =1,\dots,N$ which are defined by the numbers
$n_i$ in accordance  with the formula   (\ref{R_quant}).

Let us discuss the term $\hat{U} $  that is  responsible for the interaction  in the formula Eq. (\ref{ham_q_full}).
In general, operator $\hat{U} $ has the following form
\begin{equation}
\label{hat_U}
\hat{U} =  \sum_{m,n = 1}^\infty \varepsilon_{m,n} \hat{U}_{m\leftrightarrow n}\,,
\end{equation}
where the sequence  $\varepsilon_{m,n}$ is some finite decreasing  sequence.
The constants $\varepsilon_{m,n}$ are  the coupling constants which define  the  intensity of the vortex  interaction.
Operators  $\hat{U}_{m\leftrightarrow n}$ define the  reconnection of the considered vortex filaments in the flow.  
Each such operator describes the transformation of  $m$ vortex rings into $n$  rings and vice versa, $n \to m$.
To fix the exact form of operators  $\hat{U}_{m\leftrightarrow n}$, additional assumptions are needed.
 

For example, let us consider the operator $\hat{U}_{2\leftrightarrow 2}$  in  case of a  paired interaction between our ''structured particles''.
Thus, any two particles  located in the points with coordinates $\boldsymbol{q}_1 $ and $\boldsymbol{q}_2 $ interact by means of potential 
${\hat{\mathcal V}}(\boldsymbol{q}_1 - \boldsymbol{q}_2)$. In this case 
$$ \hat{U}_{2\leftrightarrow 2} =  \sum_{n_1,n_2,n_1^{\prime},n_2^{\prime}}\hat{U}_{2\leftrightarrow 2}(n_1,n_2,n_1^{\prime},n_2^{\prime})\,,$$
where (see Ref. \cite{AkhPel}, for instance)
\begin{eqnarray}
\label{U_2_2}
~&~&~ \hat{U}_{2\leftrightarrow 2}(n_1,n_2,n_1^{\prime},n_2^{\prime})  = \\[2mm]
~&=&~ \int\dots\int d\boldsymbol{q}_1 d\boldsymbol{q}_2 
\hat{\mathfrak a}^+(\boldsymbol{q}_1; n_1)\hat{\mathfrak a}^+(\boldsymbol{q}_2;n_2)
{\mathcal V}(\boldsymbol{q}_1 - \boldsymbol{q}_2)  \hat{\mathfrak a}(\boldsymbol{q}_2;n_2^{\prime})\hat{\mathfrak a}(\boldsymbol{q}_1;n_1^{\prime}) \,.\nonumber
\end{eqnarray}

The notations 
$$  \hat{\mathfrak a}(\boldsymbol{q}; n) = \int \hat{\mathfrak a}(\boldsymbol{p};n) e^{i\boldsymbol{q}\boldsymbol{p} } d\boldsymbol{p}\,,\qquad 
\hat{\mathfrak a}^+(\boldsymbol{q};n) = \int \hat{\mathfrak a}^+(\boldsymbol{p};n) e^{-i\boldsymbol{q}\boldsymbol{p} } d\boldsymbol{p}\,$$
were introduced here. Note that the conjugation rules
$$ \hat{U}^+_{2\leftrightarrow 2}(n_1,n_2,n_1^{\prime},n_2^{\prime})    =  \hat{U}_{2\leftrightarrow 2}(n_1^{\prime},n_2^{\prime},    n_1,n_2)$$ 
are fulfilled so  the operator   $\hat{U}_{2\leftrightarrow 2}$ is self adjoint operator.

Let us consider the $\delta$ - interaction betweeen the particles:
$$   {\mathcal V}(\boldsymbol{q}_1 - \boldsymbol{q}_2)  = \delta(\boldsymbol{q}_1 - \boldsymbol{q}_2)\,.$$
In this case, and taking into account the conservation laws for the momentum and the energy, we can write the following expression for the function
$\hat{U}_{2\leftrightarrow 2}(n_1,n_2,n_1^{\prime},n_2^{\prime})$:

\begin{eqnarray}
 ~ & ~ & ~\hat{U}_{2\leftrightarrow 2}(n_1,n_2,n_1^{\prime},n_2^{\prime})  =  \delta_{n_1+n_2,n_1^{\prime}+n_2^{\prime}}
 \int\dots\int\, d\boldsymbol{p}_1  d\boldsymbol{p}_2 d\boldsymbol{p}_1^{\prime}  d\boldsymbol{p}_2^{\prime} 
\, \times \nonumber \\
~~ & \times & \delta(\boldsymbol{p}_1 + \boldsymbol{p}_2 - \boldsymbol{p}_1^{\prime} -\boldsymbol{p}_2^{\prime})
\hat{\boldsymbol{\mathfrak a}}^+(\boldsymbol{p}_1;n_1)\hat{\boldsymbol{\mathfrak a}}^+(\boldsymbol{p}_2;n_2)
\hat{\boldsymbol{\mathfrak a}}(\boldsymbol{p}_1^{\prime};n_1^{\prime})\hat{\boldsymbol{\mathfrak a}}(\boldsymbol{p}_2^{\prime};n_2^{\prime}) \,, \nonumber
\end{eqnarray}

As it seems, some  function ${\mathcal U}_{n_1,n_2}(\boldsymbol{p}_1,  \boldsymbol{p}_2 )$
 should be added in the integrand expression  for the  reasons of convergence of the  integrals. 
For example, 
${\mathcal U} = 1$ for ${\mathcal E}(\boldsymbol{p}_1,  \boldsymbol{p}_2; n_1,n_2) \le {\mathcal E}_{max}$ 
and ${\mathcal U} = 0$ in the opposite case. The function  ${\mathcal E}(\boldsymbol{p}_1,  \boldsymbol{p}_2; n_1,n_2)$ is defined by the formula Eq. (\ref{eigen_1}).  Of course, other methods of  the  ultraviolet cutoff procedure  are also possible.

{ Taking into account the vortex nature of our structured particles, it also makes sense to consider non-local interactions of a general kind. 
This can be done, for example, by replacing in  the formula Eq. (\ref{U_2_2})

\begin{eqnarray}
~&~&~\hat{\mathfrak a}^+(\boldsymbol{q}_1; n_1)\hat{\mathfrak a}^+(\boldsymbol{q}_2;n_2)
{\mathcal V}(\boldsymbol{q}_1 - \boldsymbol{q}_2)  \hat{\mathfrak a}(\boldsymbol{q}_2;n_2^{\prime})\hat{\mathfrak a}(\boldsymbol{q}_1;n_1^{\prime}) \longrightarrow \nonumber\\[2mm]
~&~&~\int\dots\int d\boldsymbol{q}_1^{\prime} d\boldsymbol{q}_2^{\prime} 
\hat{\mathfrak a}^+(\boldsymbol{q}_1^{\prime}; n_1)\hat{\mathfrak a}^+(\boldsymbol{q}_2^{\prime};n_2)
{\mathcal F}(\boldsymbol{q}_1 - \boldsymbol{q}_1^{\prime},       \boldsymbol{q}_2 -\boldsymbol{q}_2^{\prime})
  \hat{\mathfrak a}(\boldsymbol{q}_2^{\prime};n_2^{\prime})\hat{\mathfrak a}(\boldsymbol{q}_1^{\prime};n_1^{\prime})\,,\nonumber
\end{eqnarray}

where the function $ {\mathcal F}(\cdot\, ,  \cdot)$ is certain form-factor. }

Any summands $\hat{U}_{m\leftrightarrow n}$ in the sum Eq. (\ref{hat_U}) can be constructed  similarly.
Definitely, the description of a turbulent flow in the framework  of the suggested method requires a large number of terms in the sum Eq. (\ref{hat_U}).

{ In this paper, we consider the simplest case when interacting rings retain their shape. 
Of course, the general case of interaction must take into account the change in the shape of the rings. 
The shape change can be taken into account by replacing  the $H_b$ space to the  Fock space $H_F$ , which is introduced in the  work  \cite{Tal}.
The author hopes to return to this issue in subsequent works.}

Finally, we have the  following  general expression for the partition function of the quantum turbulent flow in our model:
\begin{equation}
\label{Z_int}
  {\mathcal Z}  = {\rm Tr} \exp\left(- \frac{\hat H}{{\sf k}_B{\sf T}}\right)\,.
\end{equation}

\section{Concluding remarks}

~~~ In this paper, we have proposed the basics of the approach to the description of a quantum turbulent flow as a system of interacting vortices.
Specific calculations of  any thermodynamic   quantities with  a help of the formula Eq. (\ref{Z_int}) involve the  refinement of the model.
So,   we have to concretize the formula Eq. (\ref{hat_U}) in some way. In our opinion, the  values $\varepsilon_{m,n}$ together with the vortex concentration $V/N$ define  the living time of the single vortex.  Consequently, we can implement  Prandtl's hypothesis about the length of the mixing path  \cite{Prandtl} in a turbulent flow  within the framework of our theory.
 The complexity of describing such a phenomenon as quantum turbulence will require additional assumptions and subsequent investigations.
The author hopes to return to this issue in the future.

\end{document}